\documentclass{jjap3}

\usepackage[utf8x]{inputenc}
\usepackage{graphicx}
\usepackage{epsfig}
\usepackage{epstopdf}

\usepackage{lmodern}

\usepackage{xparse}
\usepackage[
monochrome
]{xcolor}

\title{RF Reflectometry for Readout of Charge Transition in a Physically Defined p-channel MOS Silicon Quantum Dot }
\author{Sinan Bugu$^{1}$\thanks{E-mail: bugu.s.aa@m.titech.ac.jp}, Shimpei Nishiyama$^{1}$$^{,2}$, Kimihiko Kato$^{2}$, Yongxun Liu$^{2}$, Takahiro Mori$^{2}$ and Tetsuo Kodera$^{1}$\thanks{E-mail: kodera.t.ac@m.titech.ac.jp} }
\inst{$^{1}$Tokyo Institute of Technology 2-12-1 Ookayama, Meguro-ku, Tokyo 152-8552, Japan  \\
$^{2}$Device Technology Research Institute (D-Tech), National Institute of Advanced Industrial Science and Technology (AIST), Central 2, 1-1-1 Umezono, Tsukuba, Ibaraki, 305-8568, Japan}
\abst{We have embedded a physically defined p-channel MOS (PMOS) silicon quantum dot (QD) device into an impedance transformer RC circuit. \textcolor{black}{To decrease the parasitic capacitance of the device which emerges in MOS devices that have a top gate, we fabricate a new device to reduce the device's top gate area from 400 $\mbox{$\mu$m}^2$ to 0.09  $\mbox{$\mu$m}^2$. Having a smaller top gate eliminates parasitic capacitance problem preventing the RF signal from reaching QD}. We show that we have fabricated a single QD properly, which is essential for RF single-electron transistor technique. We also analyze and improve the impedance matching condition and show that it is possible to perform readout of charge transition at 4.2 K by RF reflectometry. \textcolor{black}{This will enable } fast readout of charge and spin states.}  

\begin{document}
\maketitle

\section{Introduction}
Qubit based on silicon quantum dots (QDs) \cite{MauneNature,TakedaSciAdv} is an appropriate platform for quantum computers thanks to its long coherence time \cite{Veldhorst2015}, \textcolor{black}{scalability,} and compatibility with CMOS technology \cite{VeldhorstNatComm2017,VandersypenNJP2017}. Particularly p-channel MOS (PMOS) silicon QDs are promising for the development of a spin-based qubit system since hole spins have smaller hyperfine interactions than electron spins and can be controlled only with an electric field as they have strong spin-orbit coupling (SOC) \cite{Voisin,LiNanoLetter, MaurandNatComm}.
Charge sensing technique, where a capacitively coupled additional single QD (SQD) is required, has been used extensively to map out the charge states in QDs. While this technique is relatively easy to demonstrate since it only requires the devices to be properly fabricated, the integration time ($t_\textup{min}$), which is the minimum time to discern and characterize the states, should also be considered \cite{SchaalPRL2020}. For a readout to be effective, the measurement needs to be faster than the relaxation time of the system. In quantum computing, \textcolor{black}{ in order to perform single shot readout for quantum error correction,} the measurement time should be faster than the coherence time ($T_2^*$). \textcolor{black}{ For silicon qubits, $T_2^*$ has been reported to be several tens of $\mu$s \cite{yonedaNatureNanoTech2018}. } 

  \textcolor{black}{Charge sensing is performed at a cryogenic temperature that requires long wiring to connect the sample to the room temperature equipment. Considering the capacitance ($C$) in the cable's length, which has a typical value of 70 pF/m \cite{GautschiPiezoelectric} and the resistance of the sample $ R $ > 100 k$\Omega$, the bandwidth ($BW$) of the measurement is restricted by RC time constant and limited to a few kilohertz or less.}
 Instead of performing a voltage and current tests, namely DC measurement, a so-called RF single-electron transistor (RF-SET) overcomes the low-frequency restriction. \textcolor{black}{LC resonant tank circuit is used to measure RF waves reflected from the SET \cite{Schoelkopf} }. \textcolor{black}{This technique has been used to increase the BW of SET, quantum point contact \cite{ManoharanNanoLett,QinAPL,CassidyAPL, Noiri2020, BarthelPRB}, 
 and superconductor-insulator-normal thermometers\cite{SchmidtAPL}. It is also used to warrant  rapid readout of charge sensor \cite{Reilly_APL},}
 complex impedance measurement of circuits\cite{PeterssonNanoLett,ChorleyPRL,JungAPL,SchroerPRL,CollessPRL,GonzalezNatCom, AresPhyRevApl}, large gate two-dimensional systems \cite{TaskinenRevScieIns}, and nanomechanical resonators \cite{LaHayeNature}.

\indent  In this study, \textcolor{black} {we use SQD part of the PMOS silicon QD }device to readout charge transition via RF reflectometry. The device we fabricated is a physically defined QD, therefore, it does not need gates to form confinement potential \cite{YamaokaElectronTransport2016,YamaokaJJAP2018, Hiraoka} and it has less complexity. Since it is a p-channel device and has strong  SOC, it does not require additional structure for spin manipulation \cite{Kawakami,Muhonen}. \textcolor{black}{However, MOS devices with a large global top gate have large capacitance and this precludes RF signal from reaching QD area.} To tackle this disadvantage, \textcolor{black}{ we use a device with a small top gate}, as we explain in Section 3.1.

In the extended abstract for SSDM2020\cite{BuguSSDM2020} we presented readout of charge transition via RF reflectometry. \textcolor{black}{In this paper, we analyze and discuss impedance matching condition in detail.}
\section{Device fabrication}

\begin{figure}[t]
	\centering
\includegraphics[width=1\columnwidth]{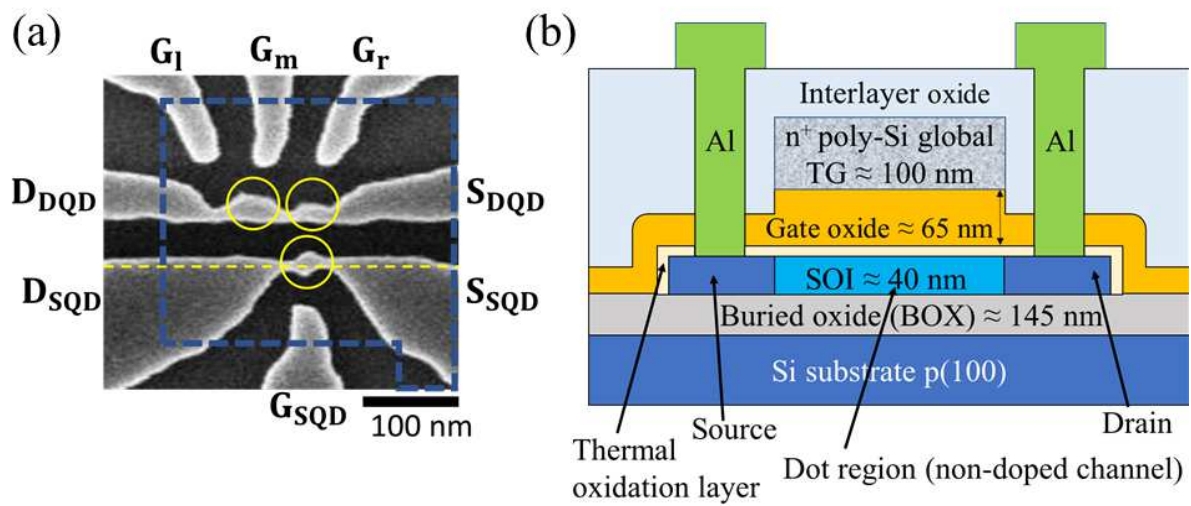}
	\caption{Physically defined p-channel silicon MOS device. (a) Scanning electron microscope image of the device taken after reactive ion etching of SOI. \textcolor{black}{ Top gate (TG) position is schematically shown by dashed blue lines.} Two QDs in DQD part and one QD in SQD part  are circled in yellow. (b) Cross-sectional schematic of the device structure along yellow dashed line in (a).}
	\label{fig:fig1ab}
\end{figure}

A scanning electron microscope (SEM) image and schematic of physically defined PMOS double QD (DQD) and SQD on undoped silicon-on-insulator (SOI) substrate are shown in Fig. 1(a), and 1(b), respectively. The dark and bright regions in Fig. \ref{fig:fig1ab}(a) indicate the BOX layer of the structure, and the SOI layer, respectively. The device has three side gates ($ \textup{G}_\textup{l} $, $ \textup{G}_\textup{m} $, $ \textup{G}_\textup{r} $) capacitively coupled to DQD and one side gate ($ \textup{G}_\textup{SQD} $) capacitively coupled to SQD for controlling electrochemical potential.  \textcolor{blue}{The device is fabricated on (100) SOI substrate. The confinement in MOS structure devices lifts the degeneracy between heavy hole state and light hole state so that the lowest sub-band corresponds to heavy holes. Therefore, heavy holes give the largest contribution to carrier properties. The in-plane effective mass of heavy hole is larger than that of electron in (100) SOI substrate, requiring stronger confinement. Hence, relatively small, $\sim$50-nm in diameter of QDs are required. The fabrication process is as follows.} A 40-nm thick SOI substrate is etched to form SQD, DQD, and side gates by reactive ion etching technique after electron beam lithography. Next, the thermal oxidation to form a 3-nm-thick $\mbox{SiO}_2$ layer and the plasma chemical vapor deposition (P-CVD) to form a 65-nm-thick $\mbox{TEOS-SiO}_2$ layer were performed. Then, a 100-nm-thick $\mbox{n}^+$-poly-Si layer was deposited by low-pressure CVD as the top gate (TG) material. In order to shrink the conventional TG area of 400 $\mbox{$\mu$m}^2$ to 0.09 $\mbox{$\mu$m}^2$, we used electron beam lithography in the TG pattern formation. After the TG formation, the $\mbox{BF}_{2}^+$ ion implantation (I/I) was performed with a dose of $\mbox{ 1.5$\times$10}^{15}$ $\mbox{cm}^{-2} $ and a tilting angle of $\mbox{7$^ \circ$ }$  at a fixed energy of 10 keV to form source-drain (SD) regions. To activate the implanted impurities, the rapid thermal annealing (RTA) was carried out at $\mbox{860 $ ^ \circ$C}$  for 2 s. Finally, contact holes and aluminum electrodes were formed, and the wafers were sintered in forming gas ambient at  $\mbox{450 $ ^ \circ$C}$  for 30 min.  \textcolor{black}{In this measurement, we use only an SQD for studying the effect of reducing TG area of the device on RF reflectometry. In the future experiments, we measure spin states in the DQD by using the SQD as RF-SET. }

\begin{figure}[t]
	\centering
	\includegraphics[width=0.6\columnwidth]{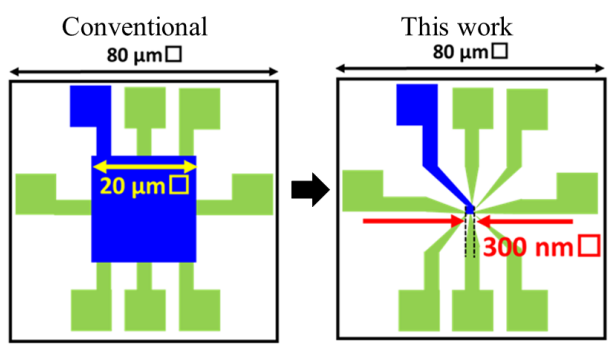}
	\caption{Schematic of the region around QD with a square TG that has a side of 20 $\mu$m (left) and 300 nm (right). Blue and green regions denote TG and SOI, respectively. To surpass large gate capacitance problem \textcolor{black}{ TG area is reduced from 400 $\mbox{$\mu$m}^2$ to 0.09 $\mbox{$\mu$m}^2$.}}
	\label{fig:fig2}
\end{figure}

\begin{figure}[t]
	\centering
	\includegraphics[width=0.6\columnwidth]{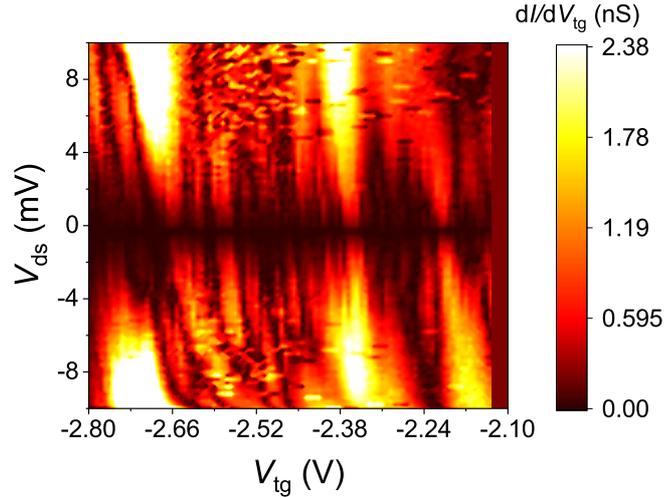}
	\caption{Charge stability diagram. TG voltage, $  V_\textup{tg} $, and drain to source voltage, $ V_\textup{ds} $, are swept to obtain Coulomb diamonds. Current can flow out of the diamonds whereas it is blocked in the diamond \textcolor{black}{ where the first derivative of the current, $  dI $/$ dV_\textup{tg} $, is minimum.} }
	\label{fig:fig3}
\end{figure}

\begin{figure}[t]
	\centering
	\includegraphics[width=0.8\columnwidth]{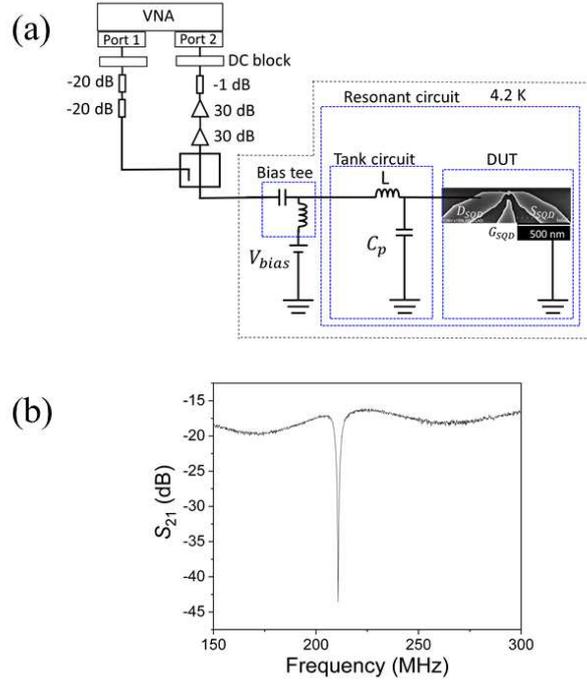} \\
	\caption{(a) Measurement setup. Attenuated RF signal is sent through series LC resonant tank circuit to drain of single quantum dot (SQD) and reflected signal that is separated by directional coupler is amplified at room temperature before it reaches the second port of vectoral network analyzer. (b) $S_\textup{21}$ as a function of frequency. Resonant frequency, $ f_r $ = 210.875 MHz, of the sample is shown when TG voltage is not applied. }\label{fig:fig4}
\end{figure}

\begin{figure}[t]
	\centering
	\includegraphics[width=0.8\columnwidth]{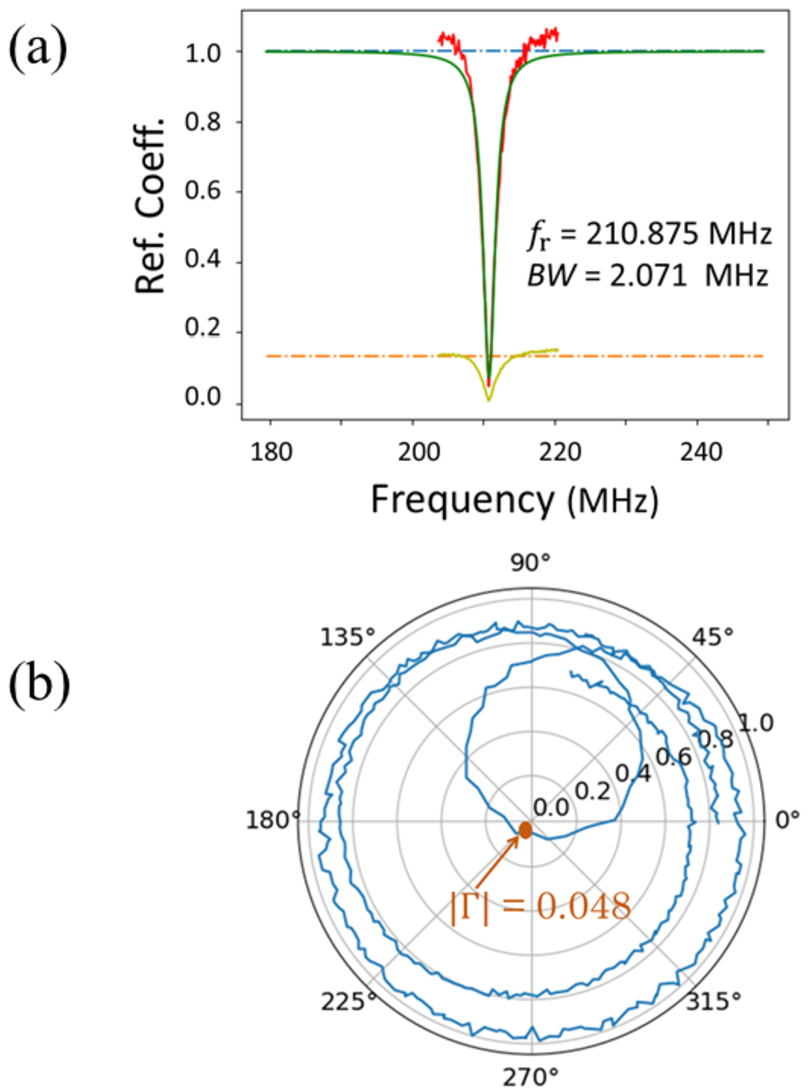} \\
	\caption{Analysis of impedance matching. (a) Background noise is subtracted (red line) from \textcolor{black}{$S_\textup{21}$ vs frequency} data shown in Fig. 4(b) and then fitted (green line) \textcolor{black}{with} a Lorentzian shape. The center frequency is 210.875 MHz and $BW$ of 3-dB regime (yellow line) is 2.0718 MHz. Loaded quality factor, $Q_\textup{L}$, is found to be 101. (b) Trace of reflected signal (blue line) is shown on a polar plot. Reflection coefficient, $|\Gamma|$, is found to be 0.048. The perfect matching occurs where  $\Gamma$ crosses the origin of the polar plot ($\Gamma$ = 0). }\label{fig:fig5}
\end{figure}

\section{Method and discussion}

\subsection{Experimental method}

In this study, we use the RF reflectometry technique to readout charge transition in a physically defined PMOS silicon QD device. \textcolor{black}{MOS devices with TG can have large gate capacitance $C_\textup{g}$, which also causes a large parasitic capacitance. Large gate capacitance prevents probing signal from reaching the QD. To suppress large gate capacitance, the TG area is reduced down from 400 $\mbox{$\mu$m}^2$ to 0.09  $\mbox{$\mu$m}^2$ as shown in Fig. \ref{fig:fig2}. }

 \textcolor{black}{We perform a DC measurement to check device characteristics where we can see single hole transition by sweeping TG voltage and drain to source voltage of SQD at a temperature of 4.2 K. We obtain multiple Coulomb diamonds that are not in the same size as shown in Fig. \ref{fig:fig3}.}
  Our measurement proves that the device we fabricated has \textcolor{black}{ a QD characteristic although some unintended localized states exist \textcolor{blue}{both in series and in parallel \cite{KoderaJJAP,Ferrus2011}. In an ideal SQD with no unintended localized states, Coulomb diamonds touch at a single point at $V_\textup{ds}=0.$ When there are localized states that charge up or get discharged, some diamonds can be shifted and overlapped. Although there exists some unintended localized states near SQD, the necessary condition of having SQD is satisfied, which is essential for studying RF-SET technique. In the RF measurement, by using the setup} shown in Fig. 4(a), we first find the resonant frequency of the device. Here, the device is mounted on an FR4 printed circuit board (PCB), and the bonding wire is connected to the drain lead of the SQD. }Before the signal reaches the sample, it is attenuated with a \textcolor{black}{ -40-dB} attenuator \textcolor{black}{to protect the device against electrostatic discharges}. Electron transition through the SQD causes a change in amplitude of the reflected signal. The directional coupler sends the carrier signal down one line, and the reflected signal up another line. The reflected signal is transmitted through the directional coupler back up the carrier line. Afterwards, the signal is amplified with room temperature amplifiers, which have a gain of 60-dB in total. \textcolor{black}{ A 2.2 $ \mu $H commercial inductor is used in a series LC resonant tank circuit shown in Fig. \ref{fig:fig4}(a). Fig. 4(b) shows the amplitude change in reflected signal ($S_\textup{21}$) as a function of applied frequency. We find a resonant dip at 210.875 MHz. From the resonant frequency $ f_\textup{r} $ of 210.875 MHz, we derive the parasitic capacitance $ C_\textup{p} $ to be 0.258 pF.}

\textcolor{black}{
\noindent The obtained value of 0.258 pF, is  lower than that obtained for the device with 400 $\mbox{$\mu$m}^2$ TG area, which was 0.6 pF \cite{YamaokaJJAP}, since the presented one has a smaller TG area.  It is worth noting that the parasitic capacitance is affected by material used in the PCB and the thickness of PCB. Besides, the distance between device and the PCB path and the number of bonding wires used to connect the device to the PCB also change the conditions. Furthermore, in actual MOSFET devices, capacitance components are distributed other than top gate capacitance. As a result, there may always not be a direct ratio between reducing top gate area and the obtained parasitic capacitance.}

\subsection{Analysis of impedance matching }
To find out how good impedance matching we analyze the matching conditions. We first subtract the background noise \textcolor{black}{ and then fit} a Lorentzian shape  (see Fig. \ref{fig:fig5}(a)). \textcolor{black}{The loaded quality factor ($Q_{\textup{L}}$) is found to be 101 by using $Q_\textup{L} = f_\textup{r}/  BW$, where 3-dB $BW$ is 2.071 MHz.} 
  We then analyze the matching on the polar plot. In Fig. \ref{fig:fig5}(b), the blue line shows the whole trace of the reflected signal. \textcolor{black}{ The reflection coefficient is given by $|\Gamma| = |{(Z_\textup{L} - Z_\textup{0})}/{(Z_\textup{L} + Z_\textup{0} )}|$, where $ Z_\textup{0} $ is the characteristic impedance of the transmission line and has a typical value of 50 $\Omega$ and $ Z_\textup{L} $ is loaded impedance. $|\Gamma|$ is found to be 0.048. Perfect matching occurs
 where $\Gamma$ crosses the origin of the polar axis ($\Gamma$ = 0). At the perfect matching point, $ Z_\textup{L} $ is equal to $ Z_\textup{0} $ \cite{Pozar}.}

\noindent\textcolor{black}{Having a $|\Gamma| $ of 0.048, our impedance matching is close to the perfect matching point.} It is worth mentioning  that one could have a better impedance matching by doing some more circuit engineering but the present matching condition is enough to perform RF-SET.

\subsection{RF readout measurement}

\begin{figure}[t]
	\centering
	\includegraphics[width=0.8\columnwidth]{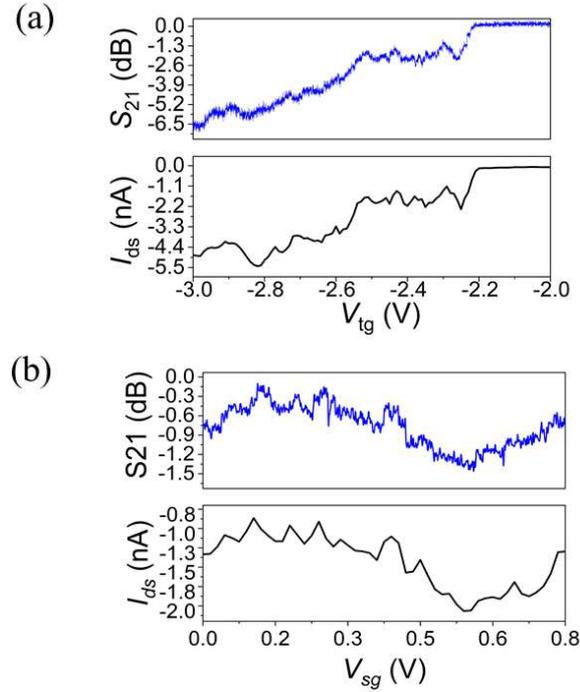} \\
	\caption{Simultaneous measurements of \textcolor{black}{ RF (shown by blue line) and DC (shown by black line) by sweeping $V_\textup{tg}$ (a) and $V_\textup{sg}$ with constant  $V_\textup{tg}$ of -2.5 V (b), respectively.  } RF and DC measurements are alike, implying that electron transition can be readout by RF reflectometry. }\label{fig:fig6}
\end{figure}

 Amplitude change in reflected signal ($S_\textup{21}$) caused by charge transition between QD and the leads is mapped out by sweeping the TG voltage as shown in the upper panel of Fig. \ref{fig:fig6}(a). \textcolor{black}{ In RF reflectometry, the frequency of probing signal should be adjusted so that it exceeds background noise \cite{Schoelkopf}. Here, we fix the RF frequency at 210.875 MHz. } At the same time, we perform standard (DC) I-V measurement, as shown in the lower panel of Fig. \ref{fig:fig6}(a). We also perform a similar measurement \textcolor{blue}{by sweeping the voltage $V_\textup{sg}$ applied to $\textup{G}_\textup{SQD} $}  while applying a constant negative voltage to the TG to form a channel between source and drain, as shown in Fig. \ref{fig:fig6}(b). Coulomb peak characteristics are similar in both DC and RF measurements, meaning that we succeeded in the readout of charge transition via RF-SET technique.

\section{Conclusions}

We have fabricated physically-defined PMOS QDs with smaller TG area and performed RF-SET measurements. Reducing TG area from 400 $\mbox{$\mu$m}^2$ to 0.09 $\mbox{$\mu$m}^2$  \textcolor{black}{resulted in surpassing the large gate capacitance problem and permitting the RF signal to reach QD area. We observed  Coulomb peaks via RF, and this result is in good agreement with DC measurement.} This result shows us a route to fast readout in our future work.
\acknowledgment
S.B. thanks to M. Fernando Gonzalez-Zalba and David J. Ibberson for fruitful discussion. The part of this work was financially supported by JST CREST (JPMJ CR1675), JSPS KAKENHI (Grant Numbers JP18K18996 and JP20H00237), and MEXT Quantum Leap Flagship Program (Q-LEAP) Grant Number JPMXS0118069228.

\end{document}